\documentclass{svproc}
\usepackage{url}

\usepackage{graphicx}
\usepackage{enumerate}
\usepackage{csquotes}

\begin{document}
\mainmatter              
\title{Precision measurements of the CKM parameters (mainly $\gamma/\phi_{3}$ measurements)}
\titlerunning{Precision measurements of the CKM parameters}  
%
\author{Prasanth Krishnan \\ (on behalf of the Belle~II Collaboration)}
\authorrunning{Prasanth Krishnan} 
%
%
\institute{Tata Institute of Fundamental Research, Mumbai, India\\
\email{prasanth.k@tifr.res.in}
}

\maketitle              

\begin{abstract}
The CKM angle $\gamma/\phi_{3}$ is the only one that is accessible with tree level decays in a theoretically clean way such that it provides a precision test of $CP$ violation in the standard model. The Belle~II experiment is a substantial upgrade of the Belle detector and will operate at the SuperKEKB asymmetric-energy $e^{+}e^{-}$ collider. The accelerator has already successfully completed the first phase of commissioning in 2016 and first $e^{+}e^{-}$ collisions in Belle~II happened during April 2018. The design luminosity of SuperKEKB is 8 $\times$ 10$^{35}$ cm$^{-2}$s$^{-1}$ and the Belle~II experiment aims to record 50~ab$^{-1}$ of data, a factor of 50 more than its predecessor~(Belle). The key method to measure $\phi_{3}$ is through interference between the $B^- \to D^0 K^-$ and $B^- \to \bar{D^{0}} K^{-}$ decays which occurs if the final state of the charm-meson decay is accessible to both the $D^0$ and $\bar{D^{0}}$ mesons. To achieve the best sensitivity, a large variety of $D$ and $B$ decay modes are required, which is possible at the Belle~II experiment as almost any final state can be reconstructed including those with photons. With the ultimate Belle II data sample of 50~ab$^{-1}$, a determination of $\phi_{3}$ with a precision of $1^{\circ}$ or better is foreseen. We explain herein the details of the planned measurements at Belle II. 
\end{abstract}
\section{Introduction}

At present, the current uncertainty on $\gamma/\phi_{3}$  is approximately 5$^{\circ}$, still worse by a factor 10 with respect to $\phi_{1} = (21.9 \pm 0.7)^{\circ}$~\cite{CKM}. From CKMfitter~\cite{CKM}, we find that the uncertainties on the CKM parameters measured from tree-level processes are larger than those from measurements of loop level diagrams. Thus, it is important to reduce the error on $\phi_{3}$ to test the validity of the standard model (SM). One of the reasons for this is the relative small branching fraction of the decays involved in the measurement owing to non-diagonal CKM matrix elements, since 

\begin{equation}
					\phi_{3} \equiv \rm{arg}\left(- \frac{V_{ud}V_{ub}^{*}} {V_{cd}V_{cb}^{*}}\right),
\end{equation}
where $V_{ij}$ is the weak vertex factor for a quark transition $i \to j$. Thus, with more data, we can improve the precision of the measurement, with the uncertainty being dominated by the available statistics.

\begin{figure}
\centering
\includegraphics[width = 0.7\columnwidth]{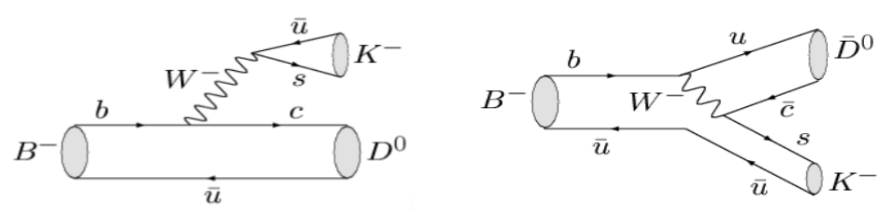}  \caption{Color-favored (left) and -suppressed (right) $B^{-} \to DK^{-}$ processes.} \label{fig:diagram}
\end{figure}

The angle $\phi_{3}$ can be extracted via interference between the color-favored $B^{-} \to D^{0} K^{-}$ and color-suppressed $B^{-} \to \bar{D^{0}}K^{-}$ decays that are shown in Fig.~\ref{fig:diagram}. These are  pure tree-level processes, hence theoretically clean. The correction to these processes is only $\mathcal{O}(10^{-7})$~\cite{PHI3CORRECTION}. If the amplitude for the color-favored decay is  $A$, then for the color-suppressed one, it is $Ar_{B}e^{i(\delta_{B} - \phi_{3})}$, where $\delta_{B}$ is the strong phase difference between the decay processes, and 
\begin{equation}
r_{B} = \frac{\mid A_{\rm{sup}} \mid} {\mid A_{\rm{fav}}\mid}.
\end{equation}
For $B \to DK$ decays, $r_{B} \sim 0.1$, whereas for $B\to D\pi$, $r_{B} \sim 0.05$. Though $B \to D \pi$ decays are not very sensitive to $r_{B}$ and $\phi_{3}$, we can use them as the control sample for $B \to DK$ to eliminate most of the systematic uncertainties.

The remainder of this document is structured as follows: Sec.~\ref{sec:methods} describes the methods and constraints to extract $\phi_{3}$, Sec.~\ref{sec:belle2} describes the potential $\phi_{3}$ sensitivity from the Belle~II experiment, as well as preliminary results from the Belle~II phase 2 data. Section~\ref{sec:summary} gives the summary.

\section{Methods for $\phi_{3}$ extraction}
 \label{sec:methods}

We classify the methods used to extract $\phi_3$ according to the $D$ meson final state:
\begin{enumerate}[(i)]
\item {\bf GLW} \cite{GLW} method: $CP$ eigenstates such as $K^{+}K^{-}$, $\pi^{+}\pi^{-}$, $K_{S}^{0}\pi^{0}$,
\item {\bf ADS} \cite{ADS} method: doubly-Cabibbo-suppressed states such as $K^{+}X^{-}$, where $X^{-}$ can be $\pi^{-}, \pi^{-}\pi^{0}, \pi^{-}\pi^{-}\pi^{+}$, and
\item {\bf GGSZ} \cite{GGSZ} method: self-conjugate multibody states such as $K_{S}^{0}\pi^{+}\pi^{-}$, $K_{S}^{+}K^{+}K^{-}$, $K_{S}^{0}\pi^{+}\pi^{-}\pi^{0}$.
\end{enumerate}

In both GLW and ADS methods, $\phi_{3}$-sensitive parameters can be extracted by taking a ratio between the suppressed and favored decay rates and a measurement of asymmetries between them. We obtain four GLW parameters $R_{CP}^{\pm} = 1 + r_{B}^{2} \pm 2 r_{B} \delta_{B} \cos \phi_{3}$ and $A_{CP}^{\pm} =  \pm 2 r_{B} \sin \delta_{B} \sin \phi_{3}/R_{CP}^{\pm}$, and two ADS parameters $R_{\rm{ADS}} = r_{B}^{2} + r_{D}^{2} + 2 r_{B} r_{D} \cos(\delta_{B} + \delta_{D})\cos\phi_{3}$ and $A_{\rm{ADS}} = 2 r_{B}r_{D} \sin(\delta_{B} \delta_{D})\sin \phi_{3}/R_{\rm{ADS}}$ for the $\phi_{3}$ extraction. Here, $r_{D}$ and $\delta_{D}$ are the ratio of the amplitudes of the suppressed and favored $D$ decays, and the $D$ strong phase, respectively. These are external inputs from charm measurements.  

\begin{figure}
\begin{tabular}{cc}
\includegraphics[width=0.5\columnwidth]{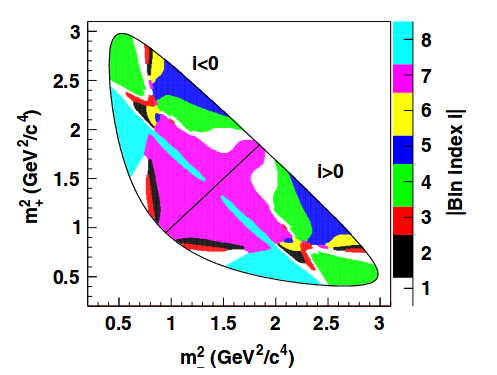} &
\includegraphics[width=0.4\columnwidth]{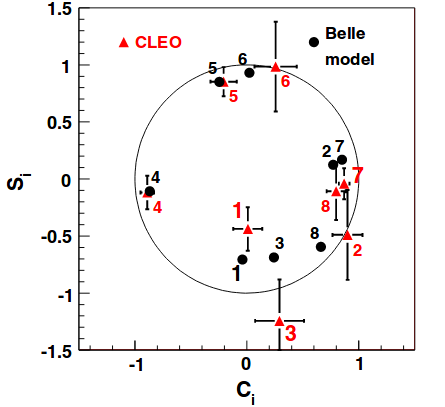} \\
\end{tabular} \caption{Binned Dalitz plot and corresponding $c_{i}$, $s_{i}$ values for $D \to K_{S}^{0}\pi^{+}\pi^{-}$ decays.}\label{fig:dalitz cisi}
\end{figure}

An inclusive approach leads to almost zero sensitivity for the GGSZ modes. Thus, we bin the Dalitz space into region with differing strong phases, which allows $\phi_{3}$ to be determined from a single channel in a model-independent manner. This eliminates the model-dependent systematic uncertainty in the measurement. Fraction of $D^{0}$ and $\bar{D^{0}}$ events in bin $i$, called $K_{i}$ and $\bar{K_{i}}$ can be  obtained from $D^{*\pm} \to D \pi^{\pm}$ decays at $B$-factories, which reduces the statistical uncertainty thanks to their large data samples. But, we need the information of these strong phases, $c_{i}$  and $s_{i}$, as external inputs from the charm factory experiments CLEO-c or BESIII, where the quantum-entangled $D^{0}\bar{D^{0}}$ pairs are produced via $e^{+}e^{-} \to \psi(3770) \to D^{0}\bar{D^{0}}$. Here, $c_{i}$ and $s_{i}$ correspond to the amplitude weighted average cosine and sine of the strong phase difference between $D^{0}$ and $\bar{D^{0}}$ decay in the $i^{\rm{th}}$ bin. An optimal binning scheme is needed to obtain the maximal sensitivity. Figure~\ref{fig:dalitz cisi} shows the Dalitz plot and the $c_{i}$ and $s_{i}$ values for the golden mode $B^{\pm} \to D(K_{S}^{0}\pi^{+}\pi^{-})K^{\pm}$ \cite{BELLEGGSZ}.

The Belle combined measurement is $\phi_{3} = (73^{+13}_{-15})^{\circ}$ \cite{CKM}, which is dominated by the GGSZ method. Similarly, the BaBar Collaboration combined all their measurements to give a value $\phi_{3} = (69^{+17}_{-16})^{\circ}$ \cite{BABAR}. The LHCb Collaboration finds $\phi_{3} = (74.0^{+5.0}_{-5.8})^{\circ}$ \cite{LHCB} by combining all their measurements. While $B$-factories used their full data sets, LHCb results are based on their Run I data. Combining these three results, we obtain $\phi_{3} = (73.5^{+4.2}_{-5.1})^{\circ}$~\cite{CKM}, which is currently dominated by the results from LHCb.

\section{Sensitivity from the Belle~II experiment}
 \label{sec:belle2}
 
The Belle~II experiment will start collecting data from early 2019 with all its sub-detectors. It will accumulate 50 ab$^{-1}$ data, about 50 times that of its predecessor, with an instantaneous luminosity of $8 \times 10^{35}~{\rm cm}^{-2} {\rm s}^{-1}$. Belle~II will also have better $K/\pi$ separation capability with the central drift chamber, imaging time-of-propagation and ring imaging Cherenkov counters, which work in different $K/\pi$ momentum ranges. An improved $K_{S}^{0}$ reconstruction efficiency is expected, mainly due to the larger acceptance of the silicon vertex detector. All this will result in a substantially improved precision measurements. More details can be found in Ref.~\cite{BELLEII}.

\begin{figure}
\centering
\includegraphics[width = 0.7\columnwidth]{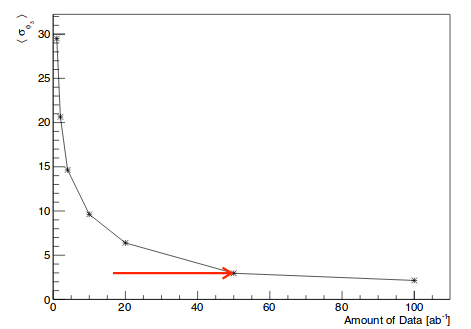}  \caption{$\phi_{3}$ sensitivity with the amount of data collected at Belle~II. The red arrow indicates the expected sensitivity from the 50~ab$^{-1}$ sample.} \label{fig:toy}
\end{figure}

Currently, the $\phi_{3}$ sensitivity is dominated by the statistical uncertainty from the number of reconstructed $B$ decays. Thus, by going from a Belle integrated luminosity of 711~fb$^{-1}$ to 50~ab$^{-1}$, the sensitivity is expected to get a significant boost. The major background is continuum events, coming from $e^{+}e^{-} \to q\bar{q}$ ($q = u, d, s, c$) due to their large cross-section. As signal hides behind this large background, we need to eliminate the latter as much as we can to get a better sensitivity. The aim is to reach a precision of $1^{\circ}$ with the full data sample, combining improvements obtained from $K/\pi$ separation, $K_{S}^{0}$ reconstruction, and continuum suppression. This would allow us to probe for possible new physics contributions, that can potentially cause a shift in the value of $\phi_{3}$ by $\pm 4^{\circ}$~\cite{1DEGPRECISION}. We perform a toy study with the golden mode $B^{\pm} \to D(K_{S}^{0}\pi^{+}\pi^{-})K^{\pm}$, which results in the expected sensitivity as a function of integrated luminosity in Fig.~\ref{fig:toy}. By adding more modes, such as $B \to D^{*}K$, $B^{\pm} \to D(K_{S}^{0}K^{+}K^{-})K^{\pm}$, we can further improve the sensitivity and approach the $1^{\circ}$ precision.

\begin{figure}
\begin{tabular}{cc}
\includegraphics[width=0.5\columnwidth]{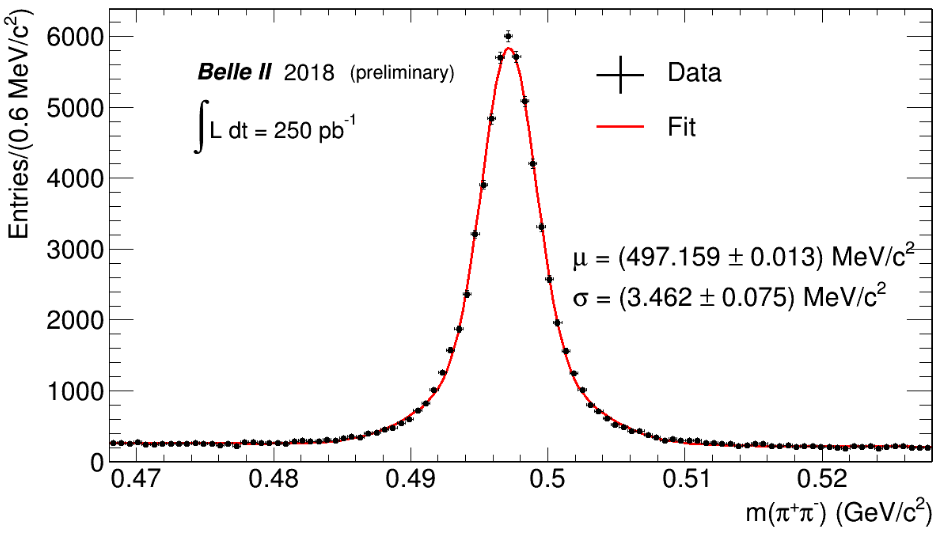} &
\includegraphics[width=0.5\columnwidth]{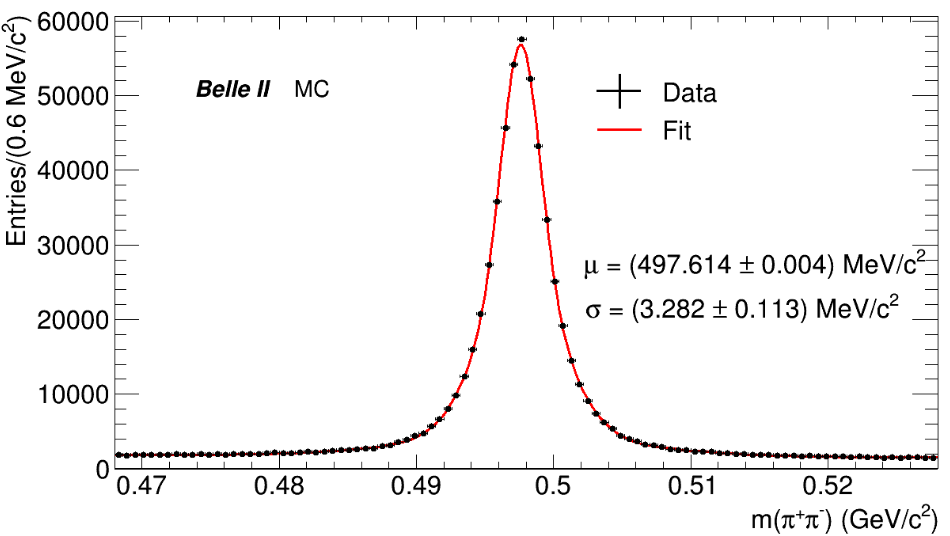} \\
\end{tabular} \caption{Comparison of $K_{S}^{0}$ reconstruction in Belle~II data and MC events.}\label{fig:phase2 ks}
\end{figure}

Recently, Belle~II successfully recorded data with all but for the vertex detector during April-May 2018. It has accumulated a data sample corresponding to an integrated luminosity of 472~pb$^{-1}$. Figure~\ref{fig:phase2 ks} shows a comparison between data and Monte Carlo (MC) simulations for the reconstruction of $K_{S}^{0}$ candidates in Belle~II. The invariant mass resolution is already showing a good agreement between data and MC events.

\begin{figure}
\begin{tabular}{cc}
\includegraphics[width=0.5\columnwidth]{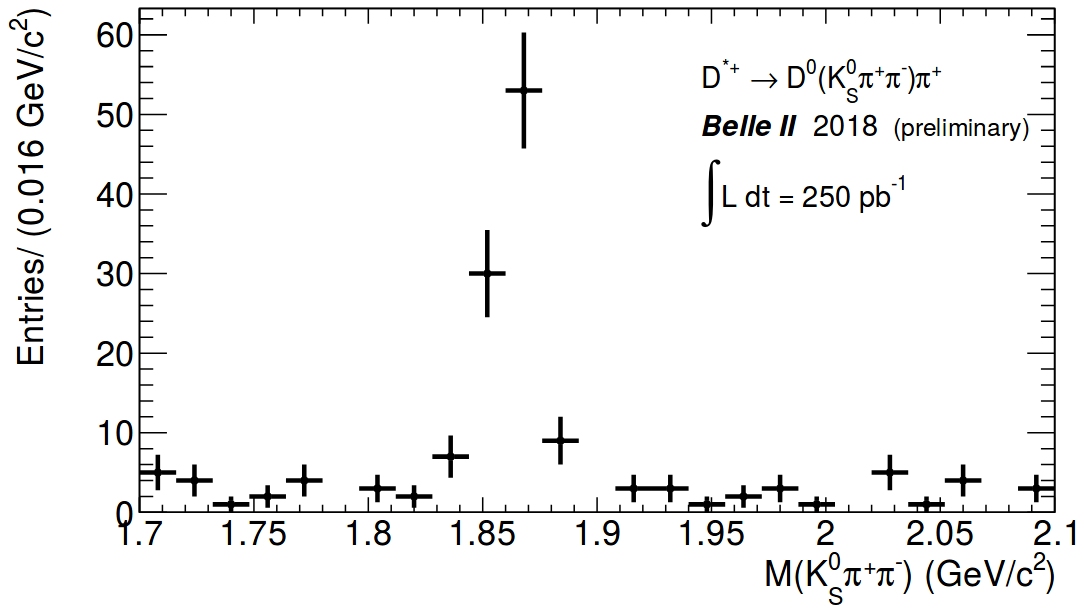} &
\includegraphics[width=0.5\columnwidth]{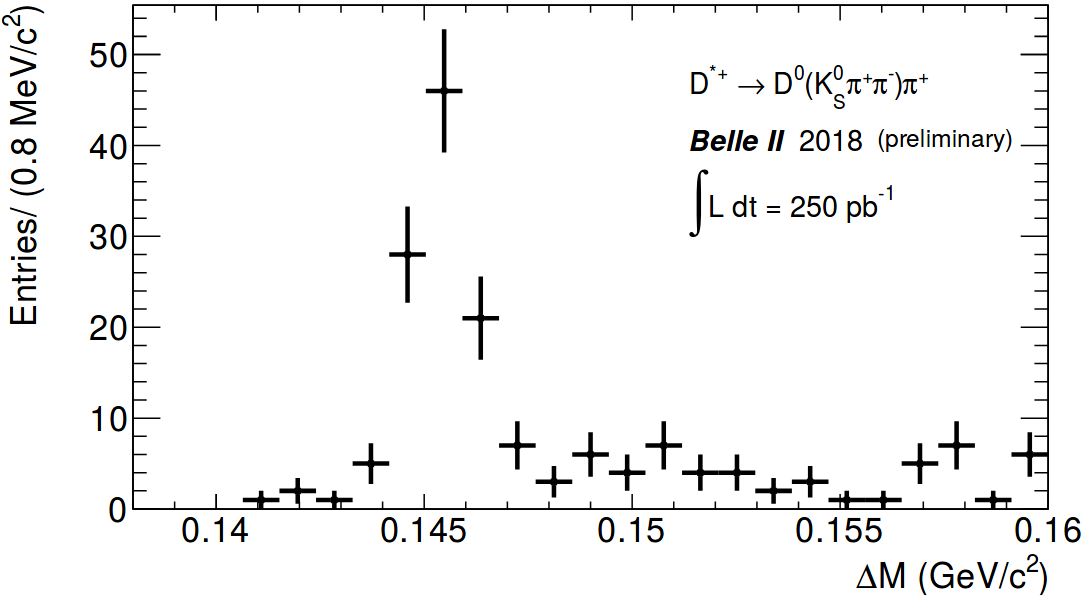} \\
\includegraphics[width=0.5\columnwidth]{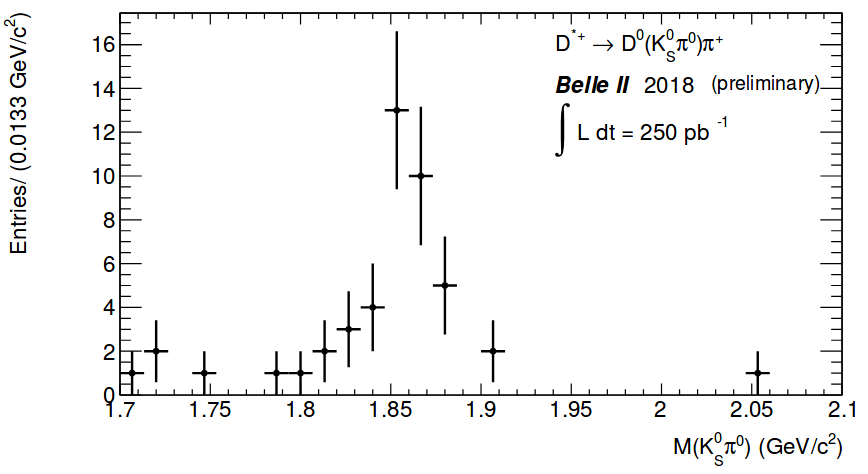} &
\includegraphics[width=0.5\columnwidth]{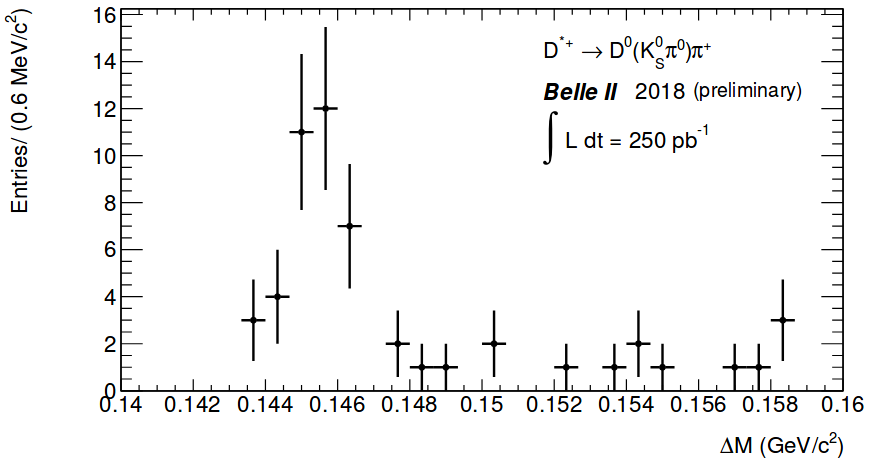} \\
\end{tabular} \caption{$M_{i}$ (left) and $\Delta M$ (right) distributions for the $D^{*}$ tagged modes $D \to K_{S}^{0}\pi^{+}\pi^{-}$ (top) and $D \to K_{S}^{0}\pi^{0}$ (bottom).}\label{fig:phase2 D*Dpi}
\end{figure}

We perform the reconstruction of $D^{*\pm} \to D\pi^{\pm}, D \to K_{S}^{0}\pi^{+}\pi^{-}$ and the $CP$ mode $D^{0} \to K_{S}^{0}\pi^{0}$, which are shown in Fig.~\ref{fig:phase2 D*Dpi}. The plotted variables are $M_{i}~(i = K_{S}^{0}\pi^{+}\pi^{-}{\rm or}~K_{S}^{0}\pi^{0}$), where $M_{i}$ is the invariant mass of the final state $i$, and $\Delta M$, the reconstructed mass difference between $M_{(i)\pi^{\pm}}$ and $M_{i}$. The signal-to-background ratio is already good in both cases, and the reconstruction of $D\to K_{S}^{0}\pi^{0}$ shows the capability of Belle~II for neutral particle reconstruction. 

\begin{figure}
\begin{tabular}{cc}
\includegraphics[width=0.5\columnwidth]{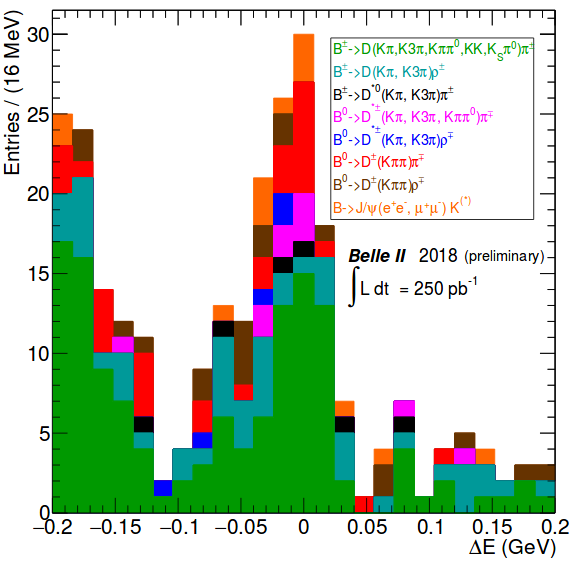} &
\includegraphics[width=0.5\columnwidth]{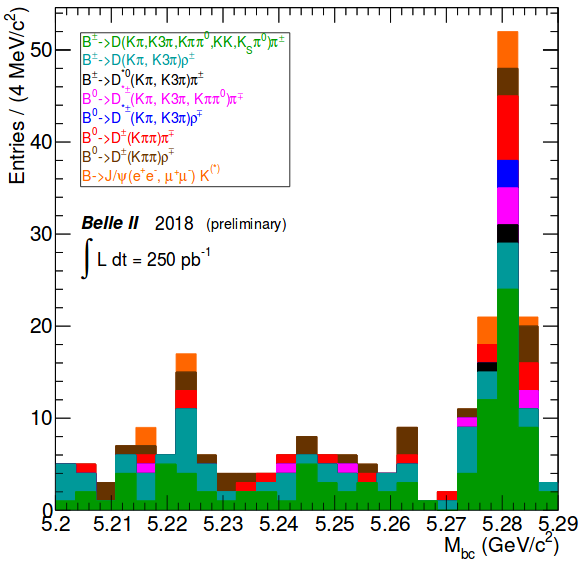} \\
\end{tabular} \caption{$\Delta E$ (left) and $M_{\rm{bc}}$ (right) distributions for various $B$ modes.}\label{fig:phase2 B}
\end{figure}

We perform the \enquote{rediscovery} of the $B$ meson from these data. We have accumulated about 100 $B$ candidates in which the majority are from the $B^{\pm} \to D\pi^{\pm}$ mode, which is the control channel for the $\phi_{3}$ extraction; corresponding distributions of fit variables are shown in Fig.~\ref{fig:phase2 B}. The variables are the energy difference $\Delta E = E_{B}^{*} - E_{\rm{beam}}^{*}$ and the beam-energy constrained mass $M_{\rm{bc}} = \sqrt{(E_{\rm{beam}}^{*}/c^{2})^{2} - (p_{B}^{*}/c)^{2}}$, where $E_{B}^{*}$ and $p_{B}^{*}$ are the energy and momentum of the  $B$ meson candidate, and $E_{\rm{beam}}^{*}$ is the beam energy, all calculated in the center-of-mass frame.

\section{Summary}
 \label{sec:summary}
 
At Belle and Belle~II, the GGSZ method will have the largest impact on the $\phi_{3}$ sensitivity. Simulation studies show that a precision of $3^{\circ}$ is achievable in Belle~II with the $D \to K_{S}^{0}\pi^{+}\pi^{-}$ mode alone, even without the full benefit particle identification, $K_{S}^{0}$ finding and continuum suppression. By combining it with all other modes and improved reconstruction algorithm, we can go closer to the goal of $1^{\circ}$ precision. Preliminary results from the recently concluded run of Belle~II without the vertex detector show promising results.


%
%

\end{document}